\documentclass{jetpl}

\usepackage[latin1]{inputenc}
\usepackage{graphicx}
\usepackage{amsmath}

\twocolumn

\lat

\title{Spin fluctuations in the stacked-triangular antiferromagnet YMnO$_3$}
\rtitle{Spin fluctuations in the stacked-triangular antiferromagnet YMnO$_3$}

\sodtitle{Spin fluctuations in the stacked-triangular antiferromagnet YMnO$_3$}

\author{B. Roessli$^1$, S.N. Gvasaliya$^{1,2}$, E. Pomjakushina$^{1,3}$ and K.
Conder$^3$}

\rauthor{B. Roessli, S.N. Gvasaliya, E. Pomjakushina and K.
Conder}
\sodauthor{B. Roessli, S.N. Gvasaliya, E. Pomjakushina and K.
Conder}
\address{$^1$Laboratory for Neutron Scattering ETHZ \& Paul Scherrer
Institute, CH-5232 Villigen PSI, Switzerland\\
$^2$Ioffe Physical Technical Institute, 26 Politekhnicheskaya, 194021,
St. Petersburg, Russia\\
$^3$Laboratory for Development and Methods, Paul Scherrer Institut, CH-5232
Villigen PSI, Switzerland
}

\dates{ 2005}{*}

\abstract{The spectrum of spin fluctuations in the stacked-triangular antiferromagnet 
YMnO$_3$ was studied above the N\'eel temperature 
using both unpolarized and polarized inelastic neutron scattering. 
We find an in-plane and an out-of-plane excitation. The in-plane mode has
two components just above T$_N$, a resolution-limited central peak and a 
Debye-like contribution. The quasi-elastic fluctuations have a line-width
that increases with $q$ like $Dq^z$ and the dynamical exponent $z$=2.3. The 
out-of-plane fluctuations have a gap at the magnetic zone center 
and do not show any appreciable $q$-dependence at small wave-vectors.
}

\begin{document}

\maketitle

\section{Introduction}
YMnO$_3$ belongs to the family of RMnO$_3$ (R=Rare-earth) manganite ferroelectric 
compounds that crystallize in the hexagonal space-group $P6_3cm$ below the 
paraelectric-ferroelectric phase transition ($\sim$ 900~K). 
In YMnO$_3$ the Mn$^{3+}$-ions form triangular layers
well separated from each other by Y-layers. Because in the ferroelectric phase
the 
lattice is distorted, the Mn-ions are slightly trimerized. The large
separation between adjacent layers suggests that YMnO$_3$ forms a good
candidate of a geometrically frustrated 2-Dimensional antiferromagnet.

The magnetic structure of hexagonal YMnO$_3$ was first investigated by
Bertaut and Mercier~\cite{bertaut} and re-investigated later in more details by 
Mu\~noz {\it et al.}~\cite{munoz}. Below T$_N\sim$70~K, the S=2 magnetic
moment of Mn-ions are arranged in a 120$\rm^o$ magnetic structure
 with the triangular layers at $z$=0 and $z$=1/2
being antiferromagnetically coupled. At saturation the magnetic moment is
$\mu$=2.9~$\mu_B$, i.e. significantly reduced from the expected 4$\mu_B$ of
Mn$^{3+}$ spins, which was taken as evidence that even in the ordered phase 
strong spin fluctuations are present as a consequence of geometrical
frustration~\cite{park}. 
Analysis of the spin-wave spectrum have confirmed the 2-Dimensional character of the magnetic 
exchange interactions in YMnO$_3$~\cite{sato} with the ratio of intra- to inter-plane
exchange interactions being of the order $\sim$2$\cdot$10$^2$. Whereas
well-defined excitations are observed below the ordering temperature, a broad
inelastic signal as well as short-range correlations between the Mn magnetic
moments within the triangular layers persist well above T$_N$~\cite{park}.

In this work we investigate the $q$ and temperature dependence of the spin
excitations in YMnO$_3$ close to the N\'eel temperature. 
We were motivated by the fact that the nature of the phase transition of 
frustrated magnets is still not completely understood and that     
the critical properties of stacked 
triangular antiferromagnets have received special attention since
Kawamura~\cite{kawamura} proposed
that the critical exponents in these systems form a new universality
class. 
Experimental confirmation of the new class of (chiral) 
exponents was found in CsMnBr$_3$ by 
unpolarized~\cite{mason} and polarized neutron scattering~\cite{plakhty}
measurements as well as in Ho~\cite{plakhty_01}. Second an anomaly in the dielectric
constant $\epsilon$ with the electric field applied in the $ab$ plane 
was found in YMnO$_3$~\cite{huang} at T$_N$. The nature of the
coupling between electric and magnetic properties in hexagonal manganites is 
a subject of intense debate~\cite{kimura}.
Clearly it is required to characterise the behavior of the  
magnetic fluctuations in the vicinity of T$_N$ to understand the  
 possible relationship with the magneto-dielectric effect in YMnO$_3$~\cite{lawes}.
%
% 
%\section{Experimental results}
%\subsection{Elastic scattering}
%
\section{Experimental}
Polycrystalline YMnO$_3$ was prepared by a solid state reaction. Starting
materials of Y$_2$O$_3$ and MnO$_2$ with 99.99$\%$ purity were mixed and grounded
and then treated at temperature 1000-1200~C in air during at least 70h with
several intermediate grindings. The phase purity of the compound was checked
with conventional x-ray diffractometer (SIEMENS D500). The powder
was hydrostatically pressed in the form of rods (8 mm in diameter and $\sim$60
mm in length). The rods were subsequently sintered at 1300~C during 30h. 
The crystal growth was carried out using an Optical Floating Zone
Furnace (FZ-T-10000-H-IV-VP-PC, Crystal System Corp., Japan) with four 1000~W
halogen lamps as a heat source. 
The growth rate was 1.5 mm/h and both rods (feeding and seeding rod) were rotated
at about 20 rpm in opposite directions to ensure liquid homogeneity.
A mixture of argon with 2$\%$ of oxygen at 5.5 bar was applied during
growing. The crystal 
has the shape of a rod with 6~mm diameter and 2~cm height and a mosaic
spread better
than~1$\rm^o$.  
\begin{figure}
\centering
\includegraphics*[scale=0.35, angle=-90]{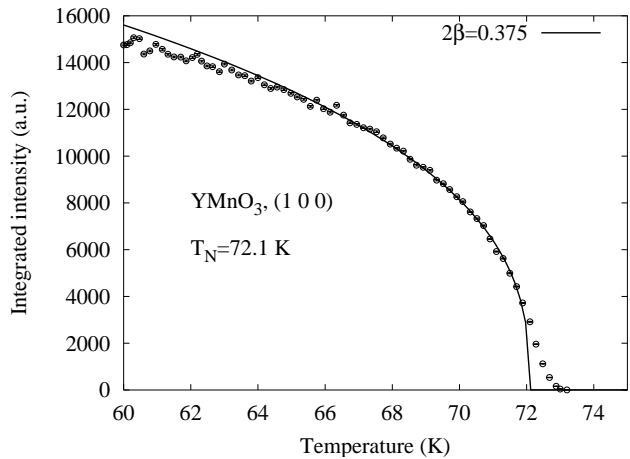}
\caption{Fig~1: Temperature dependence of the intensity of (1 0 0)
magnetic Bragg peak. The line is a fit to the data with a power law.}
\label{fig1}
\end{figure}

The measurements were performed at triple-axis
TASP located on the cold source of the neutron spallation source
SINQ. The sample was mounted inside an He-flow cryostat 
with the crystallographic axis a$^*$ and c$^*$ in the scattering plane. The
spectrometer was operated with the energy of the scattered neutrons kept
fixed at k$_f$=1.47$\rm\AA^{-1}$ for measurements with unpolarized neutrons. 
$80'$ Soller collimators 
were installed in the incident beam and before the analyzer and the detector.
With that configuration the energy resolution at zero energy transfer is 180~$\mu
eV$. To reduce both the background and contamination by higher wavelength
neutrons a cold Be-filter was installed in the scattered beam.
The inelastic polarized neutron measurements were performed at
k$_f$=1.51$\rm\AA^{-1}$ along
$(1\pm q,0,0.1)$ and at different temperatures above $T_N$. 
To perform longitudinal-polarization analysis remanent supermirror benders
\cite{semadeni01} were inserted after the monochromator and before
the analyser. The orientation of the polarization was chosen
perpendicular to the scattering plane. 
Because magnetic fluctuations
with polarization factor parallel to the neutron spin occur in the
non-spin flip channel, the non-spin-flip data (NSF) contains the 
in-plane-fluctuations and in the spin-flip channel
(SF) only out-of-plane fluctuations are present, as will be shown below.
\section{Results and Discussion}
\begin{figure}
\centering
\includegraphics*[scale=0.35, angle=-90]{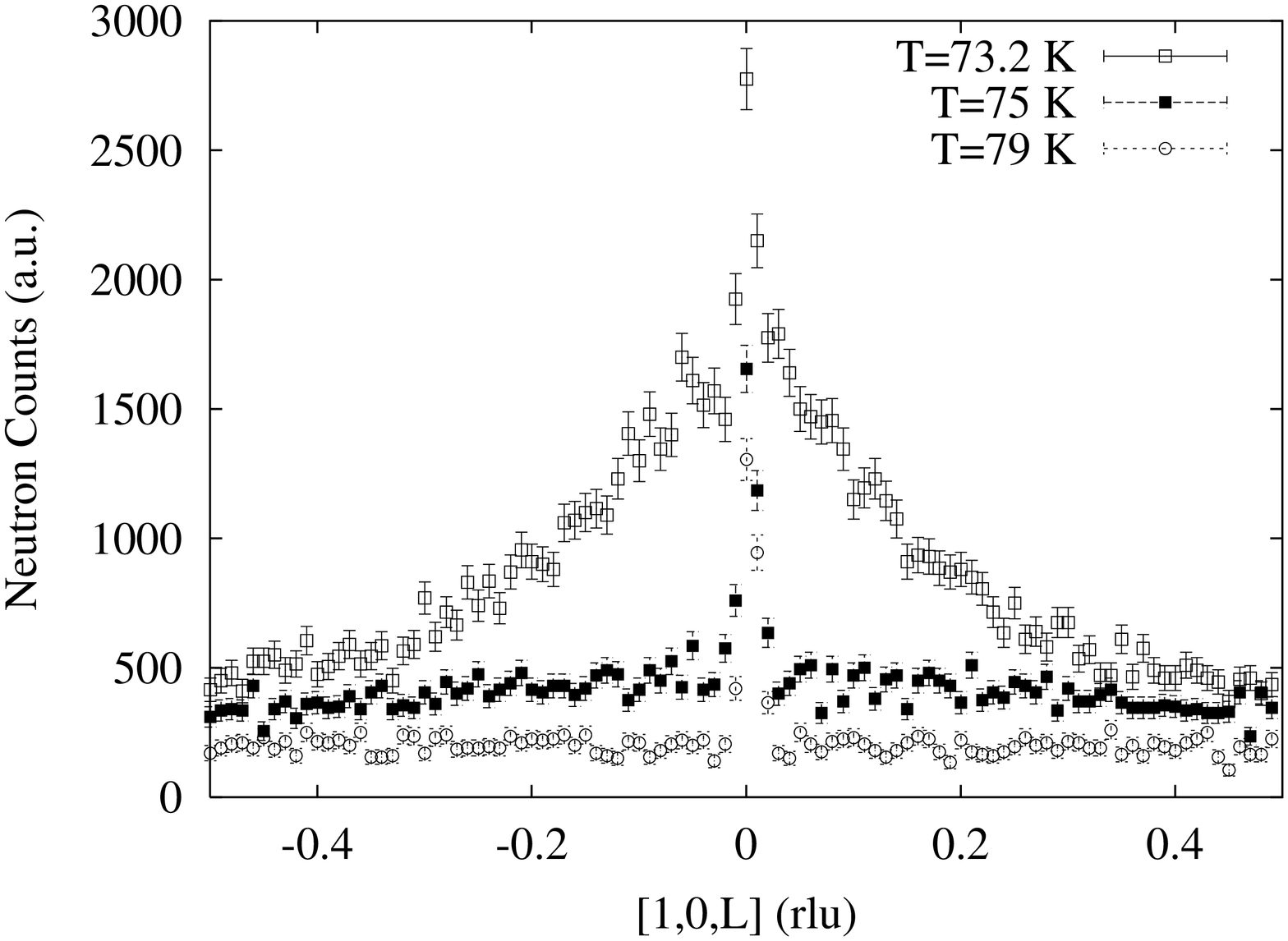}
\vspace{0.5cm}
\includegraphics*[scale=0.4]{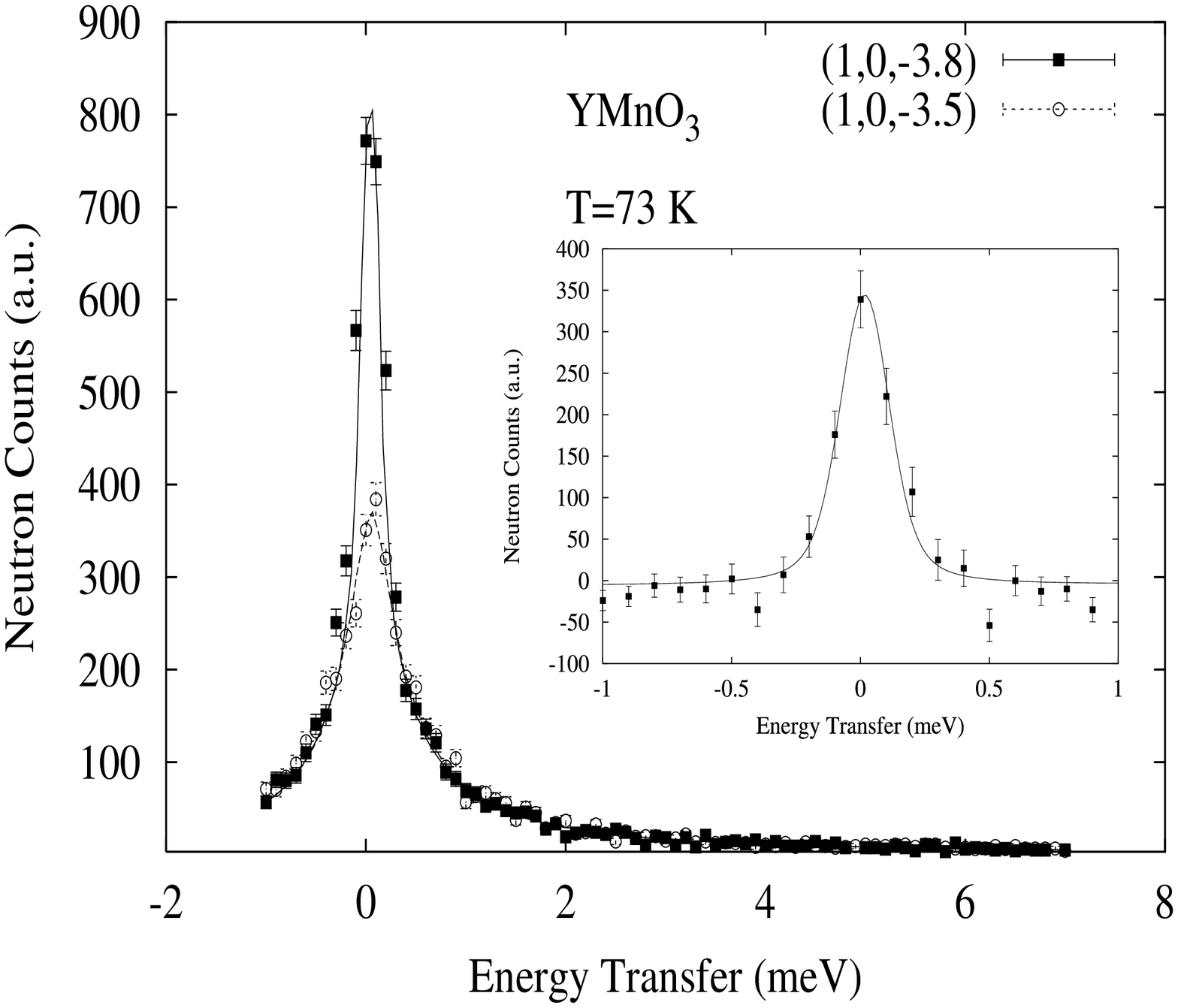}
\caption{Fig.~2: a) Elastic scans along [1,0,$l$] that show the 
increase of diffuse 
magnetic scattering in YMnO$_3$ when the temperature approaches T$_N$. b) Inelastic scan 
that show the presence of the central peak in YMnO$_3$. The insert shows the
difference between scans measured at (1,0,-3.8) and (1,0,-3.5) that
emphasizes the resolution-limited component.}
\label{fig2}
\end{figure}
Figure~\ref{fig1} shows the intensity of the (1 0 0) magnetic
Bragg reflection as a function temperature which mirrors the square
of the  
staggered magnetization. The transition temperature, as determined by
taking the derivative of the magnetization curve~\cite{bruce}, is
T$_N$=72.1$\pm 0.05$~K. The intensity of the Bragg peak follows a power law  
$I\propto |(T/T_N-1)|^{2\beta}$ with $\beta$=0.187(2). This value is lower
than reported for the case of typical stacked-triangular antiferromagnet 
RbNiCl$_3$ ($\beta$=0.28), CsNiCl$_3$ ($\beta$=0.28), CsMnBr$_3$
(0.21$<\beta<$0.25) and close to the critical exponent obtained in VCl$_2$ 
($\beta$=0.20)~\cite{kawamura,collins}. Typical two-dimensional XY  
antiferromagnets (BaNi$_2$(PO$_4$)$_2$), or ferromagnets (Rb$_2$CrCl$_4$,
K$_2$CuF$_4$) have a magnetization exponent that corresponds to the expected theoretical
value $\beta$=0.23~\cite{bramwell}. 
We note that $\beta$=0.19 in the three-dimensional triangular
Ising antiferromagnet~\cite{heinonen} 
\begin{figure}
\centering
\includegraphics*[scale=0.35, angle=-90]{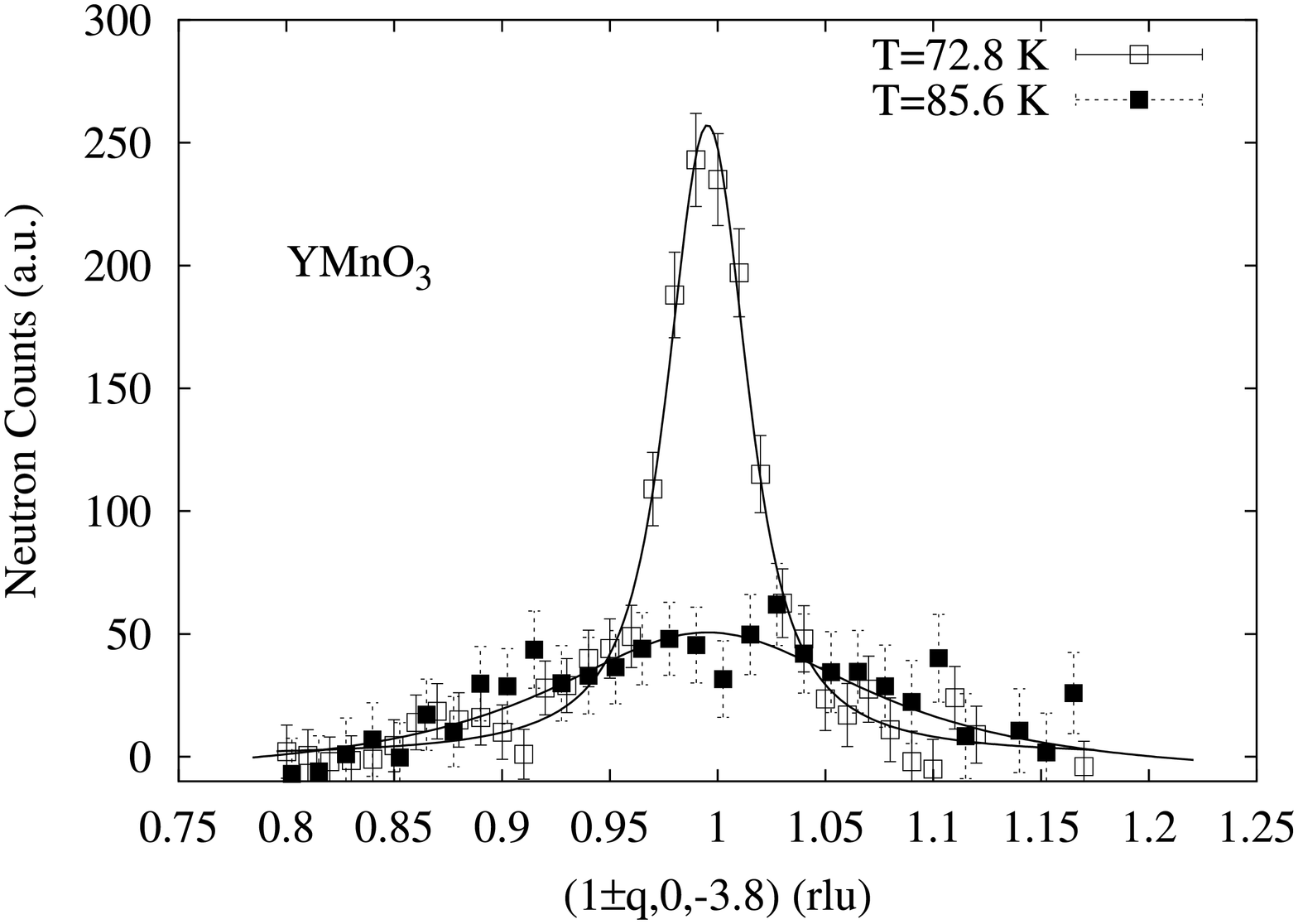}
%\vspace{1cm}
\includegraphics*[scale=0.35, angle=-90]{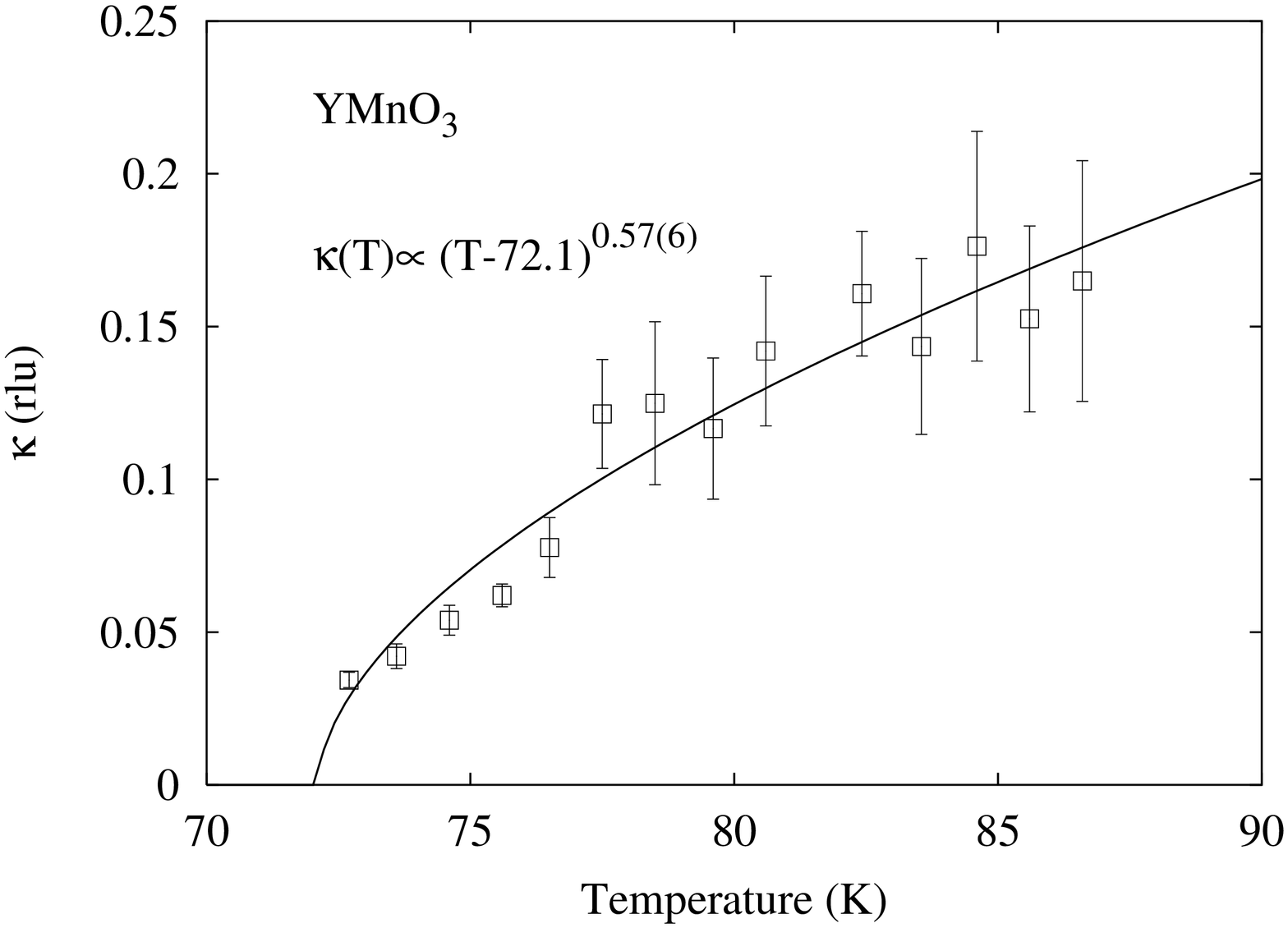}
\caption{Fig.~3: a) Neutron diffuse intensity in YMnO$_3$. The background was  
measured at T=300 K and has been subtracted. 
b) Temperature dependence of the inverse of the correlation length. The line is a fit to the data with a power law.}
\label{fig3}
\end{figure}

Figure~\ref{fig2}a shows
elastic scans along the [1 0 $l$] direction as a function of temperature. Close to the
N\'eel temperature, correlations between adjacent hexagonal planes give rise
to broad magnetic scattering along $c^*$ that eventually appear to condense 
into the magnetic Bragg peak
at T$_N$. From Fig.~\ref{fig2}b we see that this diffuse scattering is 
static on the time-scale of our experiment and corresponds to slow
fluctuations of in-plane character, as will be shown below. 
With increasing temperature these
correlations disappear and only scattering across the [1 0 $l$]-direction show
a peak in the neutron cross-section. Hence, well above T$_N$  
correlations persist only between Mn moments located 
in the hexagonal plane, as shown in Fig.~\ref{fig3}.  
Here we describe 
the line-shape of the diffuse intensity by a
Lorentzian profile convoluted with the resolution function of the
spectrometer:
\begin{equation}
\label{lor}
\chi_c(\mathbf{Q})=
{\chi_0\over{\pi}}
\frac{\kappa^2_{\parallel}}{(\mathbf{q_{\parallel}}-\mathbf{\mathbf{Q}_0})^2+\kappa^2_{\parallel}}
\frac{\kappa^2_{\perp}}{(\mathbf{q_{\perp}}-\mathbf{\mathbf{Q}_0})^2+\kappa^2_{\perp}}
\delta(\omega)
\end{equation}
\noindent where $\mathbf{Q}_0$ is the position of the magnetic rod in reciprocal space;
$\parallel$ and $\perp$ denotes direction along and perpendicular
to the magnetic rod;
$\kappa_{\parallel}$ 
the inverse of the correlation length $\xi$ between Mn-spins in the
hexagonal plane. Close to the N\'eel
temperature, the temperature dependence of $\xi$ behaves like 
$\xi (T)=0.038(\pm 0.005)\cdot (T-T_N)^\nu$ with $\nu={0.57(\pm0.06)}$ as
shown in Fig.~\ref{fig3}b.
% 
%\subsection{Inelastic unpolarized neutron spectra}
%
\begin{figure}
\vspace{0.5cm}
\centering
\includegraphics*[scale=0.35, angle=-90]{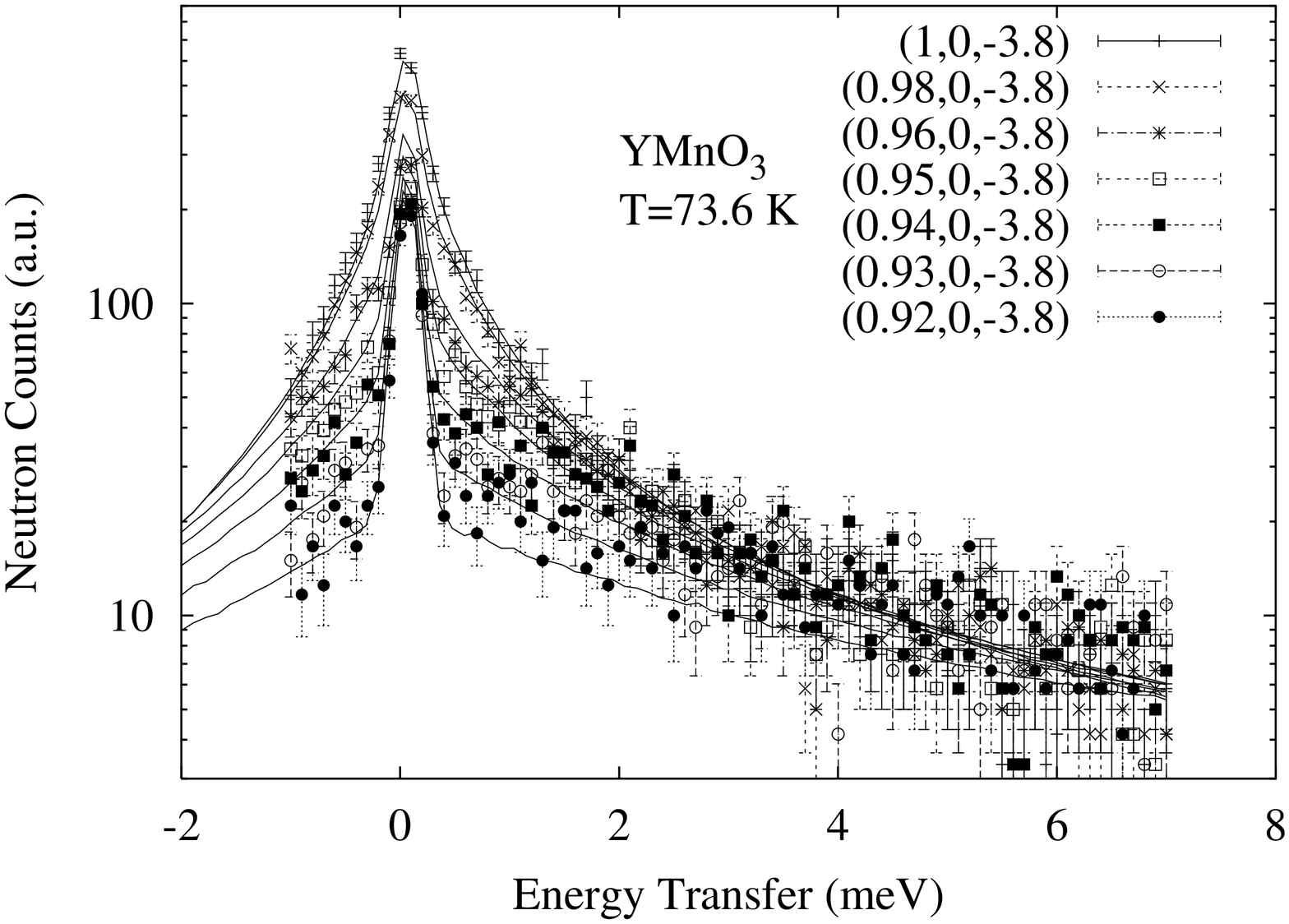}
\vspace{0.5cm}
\includegraphics*[scale=0.35, angle=-90]{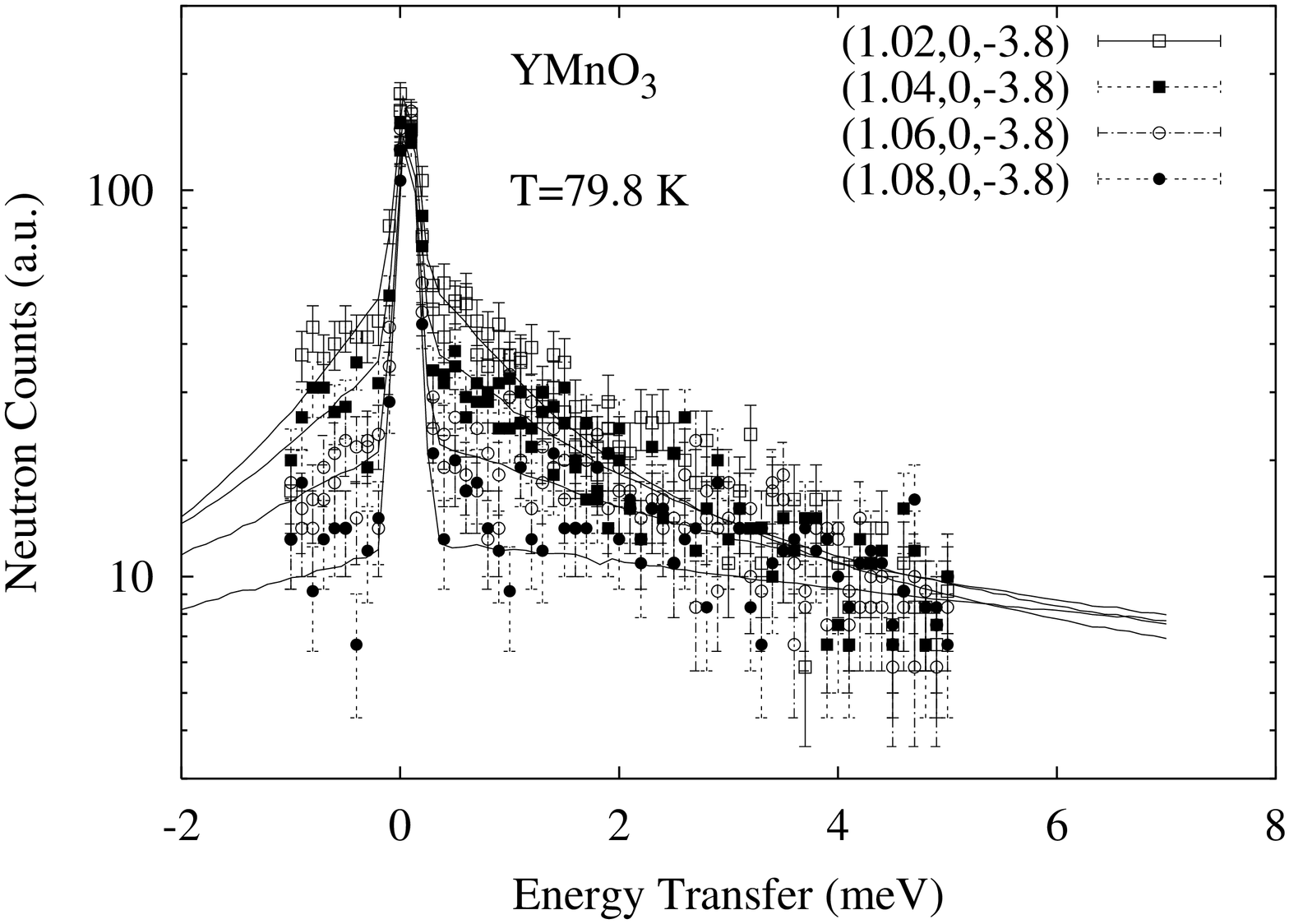}
\caption{Fig.~4: Constant q-scans in YMnO$_3$ at T=73.6~K and 79.8~K that show the in-plane 
fluctuations, as explained in the text.
The lines are the result of fit to the
data using Eq.~\ref{cs}.}
\label{fig4}
\end{figure}
We turn now to the behavior of the paramagnetic fluctuations in YMnO$_3$. 
Figure~\ref{fig4} shows typical energy scans performed at 
$\mathbf{Q}$=(1-q,0,-3.8) and T=73.6~K. 
The inelastic cross-section for an unpolarized neutron beam is given by 
 \begin{equation}\label{cross}
    \frac{d^{2}\sigma}{d\Omega d\omega}\propto\sum_{\alpha\beta}
(\delta_{\alpha\beta}-\hat{Q}_{\alpha}\hat{Q}_{\beta})S^{\alpha\beta}(\bf
Q,\omega),
 \end{equation}
where $\delta_{\alpha\beta}$ is the Kronecker symbol,
$\alpha,\beta$ are the Cartesian coordinates $x,y,z$, 
$(\bf Q,\omega)$ denote the momentum and energy transfers from
neutron to sample and $\hat{Q}={\bf Q/|Q|}$. The first term
in equation (\ref{cross}) is a selection rule that implies that only
spin components perpendicular to the scattering vector contribute
to the neutron scattering cross-section. Hence, for 
scattering vectors with large $l$-components like ${\mathbf Q}=(1\pm
q,0,-3.8)$, the inelastic spectrum 
contains essentially paramagnetic fluctuations with in-plane character
($\parallel$). 
In addition, close to the N\'eel temperature, inelastic scans through the
magnetic rod show the resolution-limited central peak described
by Eq.~\ref{lor} as shown in Fig.~\ref{fig2}. 
Thus, to analyze the data shown in Fig.~\ref{fig4},
we modeled the inelastic intensity $I(\mathbf{Q},\omega)$
in the following way
\begin{eqnarray}
I(\mathbf{Q},\omega)&=&(S_{para.}^\parallel(\mathbf{Q},\omega) \nonumber \\
                    & &+S_{inc}(\omega)+S\chi_c(\mathbf{Q}))\otimes
R(\mathbf{Q},\omega)+Bck
\label{cs}
\end{eqnarray}
\noindent where $S_{inc.}=A\delta (\omega)$ refers to the resolution-limited
incoherent
scattering that was measured at high temperature and $S$ is a scale factor. The neutron scattering function
$S_{para.}^\parallel(\mathbf{Q},\omega)$, which is related to the imaginary part of
the dynamical
susceptibility through
%the
%fluctuation-dissipation theorem
$\pi (g\mu_B)^2
S_{para.}^\parallel(\mathbf{Q},\omega)=F^2(Q)(1-\exp(-\hbar\omega/k_BT))^{-1}\Im
\chi^\parallel 
(\mathbf{Q},\omega)$, describes the line-shape of the paramagnetic
scattering as a function of momentum ($\mathbf{Q}$) and energy
($\hbar\omega$)
transfer, respectively. $F(Q)$ is the magnetic form factor of Mn. In
Eq.~\ref{cs}, the symbol $\otimes$ stands for the convolution with the
spectrometer resolution function $R(\mathbf{Q},\omega)$~\cite{popa} and
$Bck$ denotes the background level. 
We find that a Debye-like quasi-elastic line-shape for the
imaginary part of the dynamical susceptibility 
\begin{equation}
\Im \chi^\parallel (\mathbf{Q}_0+\mathbf{q}, \omega) =
\omega\chi(\mathbf{Q}_0+\mathbf{q})
{{\Gamma (\mathbf{q})}\over{\omega^2+{\Gamma(\mathbf{q})^2}}}
\label{fl}
\end{equation}
reproduces the data adequately. 
$\chi (\mathbf{q})$ is the static susceptibility as in Eq.~\ref{lor} taken relative to the antiferromagnetic
zone center ($\mathbf {Q}_0$=(1,0,0))  and $\Gamma (\mathbf q)$ is the damping of the paramagnetic
fluctuations.
Figure~\ref{fig4} shows the results of fits to the data at T=73.6~K from which
we extract that the damping of the in-plane fluctuations $\Gamma({\bf q})$
evolves like $\Gamma(0,T)$+(1450$\pm$90)$q^z$ (meV) with $\Gamma(0,73.6~{\rm K})$=0.23 $\pm
0.07$ (meV), and $z$=2.26$\pm 0.07$. 
\begin{figure}
\vspace{0.5cm}
\centering
\includegraphics*[scale=0.35, angle=-90]{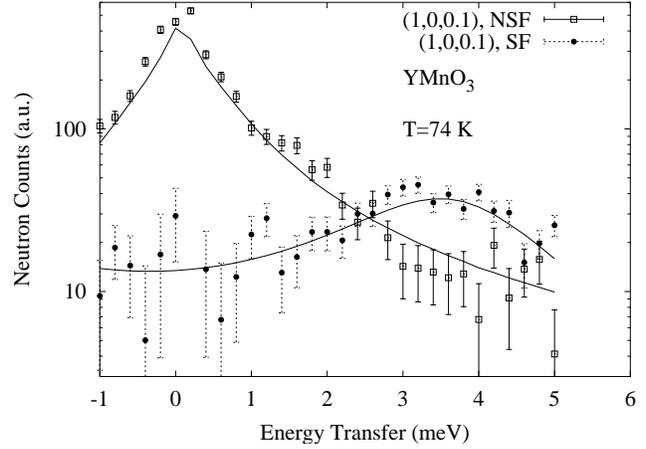}
\caption{Fig.~5: Constant q-scans in YMnO$_3$ at T=73.6~K that show the in-plane 
fluctuations in the non-spin flip (NSF) and the out-of-plane
component in the spin-flip channel (SF), respectively.
The lines are the result of fit to the
data as described in the text.}
\label{fig5}
\end{figure}
%
%\subsection{Inelastic polarised neutron spectra}
In contrast with data taken around (1$\pm q$,0,3.8) the inelastic spectra 
around (1$\pm q$,0,0.1) cannot be fitted with Eq.~\ref{fl}. This is an
indication that the spectrum of paramagnetic fluctuations in YMnO$_3$ consists
of 2 modes, an in-plane ($\parallel$) as well as an out-of-plane component
($\perp$). 
To separate the $\parallel$- from the $\perp$-fluctuations, it is necessary to
use polarization analysis. 
%To that end remanent supermirror benders were
%used to polarise the neutron beam and to analyse its polarisation after
%scattering. 
%The polarisation of the neutron beam was chosen perpendicular to
%the scattering plane. In that geometry the spin-flip channel contains the
%$\perp$-fluctuations, whereas the flucutations with $\parallel$-polarisation
%are non-spin flip. 
%
%
\begin{figure}[h]
\vspace{0.5cm}
\centering
\includegraphics*[scale=0.35, angle=-90]{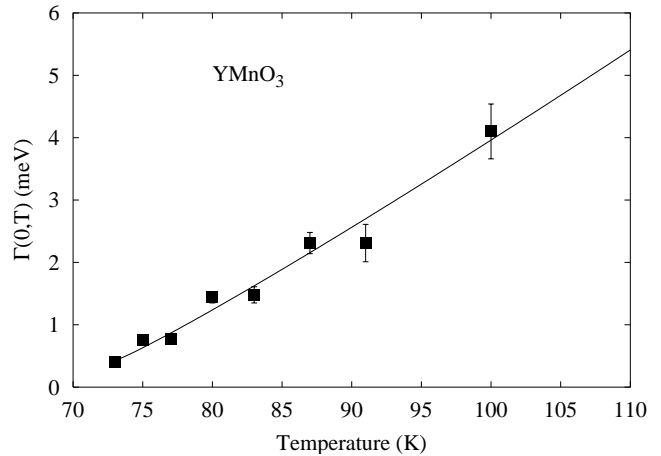}
\caption{Fig.~6: Temperature dependence of the damping of the paramagnetic 
fluctuations with in-plane polarization as measured with polarized neutrons. 
The line is a fit to the data as explained in the text.}
\label{fig6}
\end{figure}
A typical inelastic spectrum measured with polarization analysis 
is shown in Fig.~\ref{fig5} that reveals an out-of-plane
excitation in the spin-flip channel. In contrast with the $\parallel$-fluctuations, the
$\perp$-component is inelastic and is best described by a damped-harmonic
oscillator function
 \begin{equation}
 \label{dho}
  S^\perp_{DHO}\propto \frac{\omega}{1-\exp(-\omega/T)}
  \frac{\gamma_{q}}
  {(\omega^2-\Omega_{q}^2)^2+\omega^2\gamma_{q}^2}.
 \end{equation}
where $\Omega_q$ is the renormalized frequency and $\gamma_q$ the damping of
the excitation. Within the precision of the measurements we find that 
$\Omega_q$=3.9 $\pm$0.3 (meV) and $\gamma_q$=2 $\pm$0.5 (meV) in the range of momentum 
values reached in the present experiment ($q\le$0.1). 
\begin{table}[t]
        \caption{\label{damping}Table~1: Parameter values for the damping
($\Gamma(\mathbf{q})=\gamma q^z$ (meV)) and the critical exponent ($z$) of the
$\parallel$-fluctuations in YMnO$_3$.}
%         \begin{ruledtabular}
         \begin{tabular}{lccccc}
%         \hline
           &  T (K) & $\gamma$ & $\Delta \gamma$ & $z$ & $\Delta z$\\
           \hline
            unpol. &    73.6 & 1450 &  90 &  2.26 &  0.07\\
            &    79.8 & 1283 &  272 &  2.29 &  0.16\\
            pol. &    74 & 1306 &  277 &  2.53 &  0.25\\
            &    78 & 1028 &  313 &  2.40 &  0.4\\
         \end{tabular}
%         \end{ruledtabular}
\label{table1}
 \end{table}

The analysis of the data in the 
non-spin-flip channel  yields the same $q$-dependence for the
$\parallel$-fluctuations than was obtained with the unpolarized set-up. 
The results are summarized in Table~\ref{table1}. 
The mean value for the dynamical exponent
of the $\parallel$-fluctuations in YMnO$_3$ yields $z\sim$2.3 and does
not agree with the theoretical dynamical exponent $z$=1 for the classical 
2D-triangular antiferromagnet~\cite{landau}.
Also the value obtained in YMnO$_3$ is quite different from 
the dynamical exponents $z$=1.5 expected for the 3D-Heisenberg
antiferromagnet and measured e.g. in RbMnF$_3$~\cite{coldea} and
CsMnBr$_3$~\cite{mason2}.
Finally we show the temperature dependence of
$\Gamma(0,T)$ in Fig.~\ref{fig6} that increases almost 
linearly above T$_N$, $\Gamma(0,T)=0.4+0.07*(T-T_N)^{1.1\pm0.2}$.  
%The dynamic scaling relation for an antiferromagnet $\Gamma(0,T)\propto (T-T_N)^{z\nu}$  
%yields $\nu\approx$0.48(5) 
% that is in agreement with Kawamura's result for 
%layered XY-antiferromagnets $\nu$=0.54(2)~\cite{kawamura2}. 
%
\section{Conclusion}
To conclude, using both unpolarized and polarized inelastic neutron scattering, we
showed that 
there are two magnetic excitations in the paramagnetic regime of YMnO$_3$
that have in-plane and an
out-of-plane polarization, respectively. The in-plane mode has a
resolution-limited central peak and a
quasi-elastic component with a line-width  that 
increases as a function of momentum transfer $q$ like $\propto q^z$ 
with the dynamical exponent $z$=2.3. The presence of two time-scales in the
spectrum of in-plane flucutations might be the signature for
the coexistence of 2D and 3D fluctuations in the vicinity of T$_N$~\cite{regnault}. 
The
out-of-plane fluctuations are inelastic at the magnetic zone center
and do not show any $q$-dependence for small wave-vectors.%
\section{Acknowledgments}
This work has been performed at the Neutron Spallation Source SINQ,
Paul Scherrer Institut, Switzerland and was partly supported by NCCR MaNEP
project.

\end{document}